\documentclass[a4paper,11pt]{article}
\usepackage{epsfig,multicol,amsmath}
\usepackage[T1]{fontenc}
\usepackage{jheppub}
\usepackage{epstopdf}
\usepackage{amsmath}     % defines the command \dfrac (fractions in displaystyle
\usepackage{epsfig,epsf}
\usepackage{graphicx}
\usepackage{rotating}
\usepackage{verbatim}

\def\beq{\begin{equation}}
\def\eeq{\end{equation}}
\def\bea{\begin{eqnarray}}
\def\eea{\end{eqnarray}}

\def\eq#1{{Eq.~(\ref{#1})}}
\def\fig#1{{Fig.~\ref{#1}}}

\newcommand{\Lb}{\left(}
\newcommand{\Rb}{\right)}
\parskip=\itemsep               %?

\newcommand{\nn}{\nonumber}

\newcommand{\h}{\frac{1}{2}}

\newcommand{\pom}{I\!\!P}

\def\pom{{I\!\!P}}

\title{ CGC/saturation approach for soft interactions
 at high energy: survival probability of the central exclusive production}
\author[a]{E. ~Gotsman,}
\author[a,b]{   E.~ Levin}
\author[a]{  and U.~ Maor}

 \affiliation[a]{Department of Particle Physics, School of Physics and 
Astronomy,
Raymond and Beverly Sackler
 Faculty of Exact Science, Tel Aviv University, Tel Aviv, 69978, Israel}
\affiliation[b]{Departemento de F\'isica, Universidad T\'ecnica Federico Santa Mar\'ia, and Centro Cient\'ifico-\\
Tecnol\'ogico de Valpara\'iso, Avda. Espana 1680, Casilla 110-V, Valpara\'iso, Chile}
\emailAdd{gotsman@post.tau.ac.il}
\emailAdd{leving@post.tau.ac.il, eugeny.levin@usm.cl}
\emailAdd{maor@post.tau.ac.il}
\abstract{ We estimate 
the value of the survival probability 
 for  central exclusive production, in a model, which is based on the 
CGC/saturation approach. 
   Hard and  soft
 processes are described in the same framework. 
  At LHC energies,  we obtain a small value for the survival probability. The source of
 the  small value, is the impact parameter dependence of the hard 
amplitude.
  Our model  has successfully described a large body of soft
 data: elastic, inelastic and diffractive cross sections,
 inclusive production and rapidity correlations, as well as
 the $t$-dependence of deep inelastic diffractive production
 of  vector mesons.}

\keywords{CGC/saturation approach,  survival probability, soft processes }
\preprint{TAUP - 30021/15\\
\today}

\begin{document}
\maketitle
\flushbottom

\section{Introduction}
  The large body of  experimental data
  \cite{ALICE,ATLAS,CMS,TOTEM,ALICEI,ATLASI,CMSI,CMSMULT,ATLASCOR}
   on high energy soft interactions from the LHC, calls for an approach
 based on QCD   that allows us to comprehend this data. However, due to 
the
 embryonic stage of our understanding of the confinement of quarks and 
gluons,
 we are doomed to have to introduce  phenomenological model 
assumptions
 beyond that of QCD.    In our recent 
articles \cite{GLMNI,GLM2CH,GLMINCL,GLMCOR}
 we have proposed an approach, based on the CGC/saturation effective 
theory
 of high energy interactions in QCD (see Refs.\cite{GLR,MUQI,MV,B,K,MUCD,JIMWLK}
 and Ref.\cite{KOLEB} for a review)  and the 
 Good-Walker\cite{GW} approximation for the structure of hadrons.
 In the next section we  give a brief review of  our model,  however, 
  we would like to mention here, that
   the main ingredient of this model, is the BFKL Pomeron\cite{BFKL,LI},
 which describes both  hard and soft processes
 at high energies.
 In other words, in our approach, we do not separate the interactions into 
hard and soft,   both are described in the framework of the same scheme. 
The
 second important remark  concerns the description of the experimental
 data: we obtain a good description of the cross sections of elastic and
 diffractive cross sections, of inclusive productions and the rapidity
 correlations at high energies. Consequently, we feel that we are ready to
test our model on  
 a complicated phenomenon, the survival probability of
 central diffractive production.
    %%%%%%%%%%%%%%%%%%%%%%%%%%%%%%%%%%%%%%%%%%%%%%%%%%%%%%%%%%%
       \begin{figure}[ht]
    \centering
  \leavevmode
      \includegraphics[width=17cm]{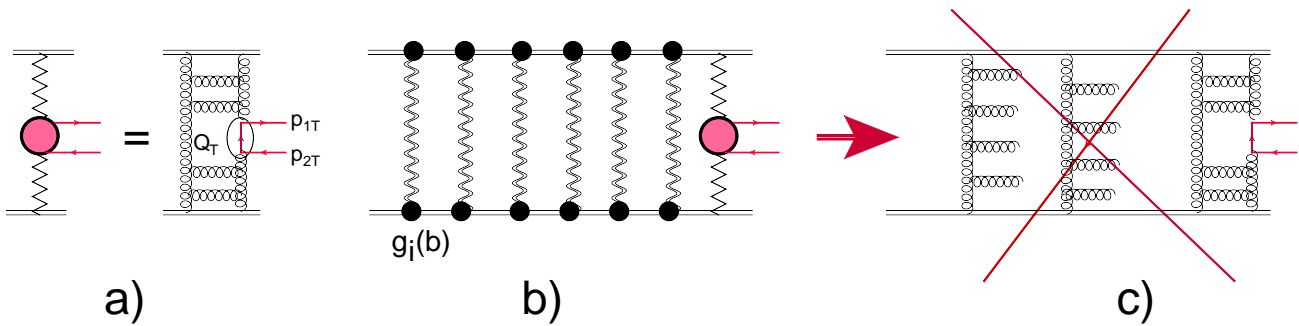}  
      \caption{\fig{sp1}-a shows the scattering amplitude of
 the hard process. \fig{sp1}-b describes       
       the set of the eikonal  diagrams in the BFKL Pomeron
 calculus, which  suppress the di-jet quark production, due to the
 contamination of the LRG (large raridity gap), by  gluons that can be 
produced by
 different parton  showers, as  shown in \fig{sp1}-c.
   }
\label{sp1}
   \end{figure}

 %%%%%%%%%%%%%%%%%%%%%%%%%%%%%%%%%%%%%%%%%%%%%  
  The physical meaning of  "survival probability" has been
 clarified in the first papers on this subject (see
 Refs.\cite{BJ,DKS,GLMSP1}),  and   is illustrated 
 by \fig{sp1}, using the example of the central production
 of a di-jet with large transverse momenta. At first sight
 we have to calculate
  the diagram of \fig{sp1}-a in  perturbative QCD, in which
 only two protons and the di-jet are produced, without any other hadrons.
    However, this   is not sufficient, since  
simultaneously, a number of parton showers can be produced,
 and gluons (quarks) from these showers will produce
 additional   hadrons.
To calculate  central diffractive production, we have to 
exclude these processes.
 In other words, we
  multiply the cross section given by the diagram of
 \fig{sp1}-a,  by the suppression factor, which reflects the
 probability of not  having any additional parton showers.
 This factor is the "survival probability".

 Even this brief description indicates, that we have a  complex
 problem, since some of the produced parton showers can have
 perturbative QCD structures, while others can stem from the
 long distances, and can be  non-perturbative by nature.
 Therefore, to attack this problem we need a model
 that describes  both long and short distances.

 As we have mentioned above, our model fulfills these requirements, and 
so 
we will proceed to 
  discuss  survival probability  in this model, expecting reliable
 results.

The next section is a brief review of our approach.  We include it in
 the paper, for the completeness of presentation, and to emphasize that
 both  short and long distance phenomena are described in the same 
framework.
 Section 3 is devoted to derivation of the formulae for the survival
 probability using the BFKL Pomeron calculus. The numerical estimates
 are given in section 4, while in the section Conclusions, we summarize 
our results.

  \section{Our model: generalities and the  elastic amplitude}
\begin{figure}[ht]
    \centering
  \leavevmode
      \includegraphics[width=14cm]{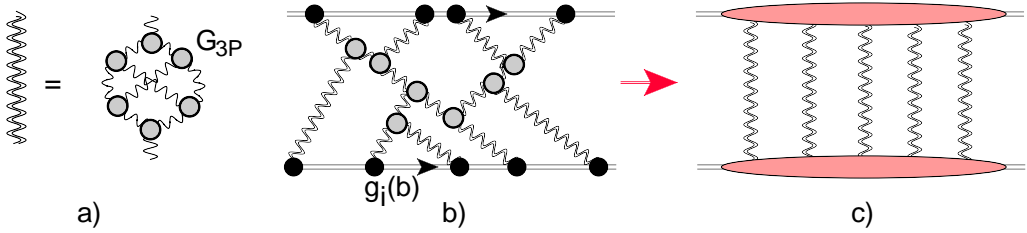}  
      \caption{\fig{amp}-a shows the set of the diagrams in the
 BFKL Pomeron calculus that produce the resulting (dressed) Green
 function of the Pomeron in the framework of high energy QCD.
 In \fig{amp}-b the net diagrams    which   include
 the interaction of the BFKL Pomerons with colliding hadrons are shown.
After integration
 over positions of $G_{3 \pom}$ in rapidity,
 the sum of the diagrams reduces to \fig{amp}-c .}
\label{amp}
   \end{figure}

  In this section we briefly  review  our model which successfully
 describes diffractive\cite{GLMNI,GLM2CH} and inclusive cross
 sections\cite{GLMINCL}. The main ingredient of our model is the BFKL
 Pomeron  Green function, which we obtained using a CGC/saturation
 approach\cite{GLMNI,LEPP}.  We determined  this function from the
 solution of the non-linear Balitsky-Kovchegov (BK) equation\cite{B,K}, 
 using the MPSI approximation \cite{MPSI} to sum enhanced diagrams, shown
 in \fig{amp}-a. It has the following form:
 \bea \label{G}
G^{\mbox{\tiny dressed}}\Lb T\Rb\,\,&=&\,\,a^2 (1 - \exp\Lb -T\Rb )
  + 2 a (1 - a)\frac{T}{1 + T} + (1 - a)^2 G\Lb T\Rb \nn\\,
~~~&\mbox{with}&~~G\Lb T\Rb = 1 - \frac{1}{T} \exp\Lb \frac{1}{T}\Rb
 \Gamma_0\Lb \frac{1}{T}\Rb.
\eea

\beq \label{T}
T\Lb s, b\Rb\,\,=\,\, \,\,\phi_0 S\Lb b, m\Rb e^{0.63\lambda \ln(s/s_0)},
~~~~\mbox{with}~~~S\Lb b , m \Rb \,\,=\,\,\frac{m^2}{2 \pi} e^{ - m b}.
\eeq 
 In the above formulae $a=0.65$, this value was chosen so as to attain the
 analytical form of the solution of the BK equation. Parameters
 $\lambda$ and $\phi_0$ can be estimated in the leading order of 
 QCD, but due to the large next-to-leading order corrections, we
 consider them as objects to be determined from  a fit to the relevant 
experimental data. $m$ is a non-perturbative
 parameter, which characterizes the large impact parameter behavior of
 the saturation momentum, as well, as the typical size of dipoles that
 take part in the interaction. The value of $m =5.25\,GeV$ in our
 model, supports our main assumption, that the BFKL Pomeron calculus,
 based on a  perturbative QCD approach, is able to describe soft 
physics,
 since $m \,\gg\,\mu_{soft}$ ,where $\mu_{soft}$ is the natural scale for
 soft processes ($ \mu_{soft} \,\sim\,\Lambda_{QCD}$ and/or  pion mass).
 
 Unfortunately, in the situation where the confinement problem is still 
far 
 from being solved, we need to rely on a phenomenological approach  for 
the 
structure of
 the colliding hadrons.
 We use a two channel model, which  allows us also to calculate the
 diffractive production in the region of small masses.
   In this model, we replace the rich structure of the
 diffractively produced states, by the single  state with the wave function 
$\psi_D$.
  The observed physical 
hadronic and diffractive states are written in the form 
\beq \label{MF1}
\psi_h\,=\,\alpha\,\psi_1+\beta\,\psi_2\,;\,\,\,\,\,\,\,\,\,\,
\psi_D\,=\,-\beta\,\psi_1+\alpha \,\psi_2;~~~~~~~~~
\mbox{where}~~~~~~~ \alpha^2+\beta^2\,=\,1;
\eeq 

Functions $\psi_1$ and $\psi_2$  form a  
complete set of orthogonal
functions $\{ \psi_i \}$ which diagonalize the
interaction matrix ${\bf T}$
\beq \label{GT1}
A^{i'k'}_{i,k}=<\psi_i\,\psi_k|\mathbf{T}|\psi_{i'}\,\psi_{k'}>=
A_{i,k}\,\delta_{i,i'}\,\delta_{k,k'}.
\eeq
The unitarity constraints take  the form
\beq \label{UNIT}
2\,\mbox{Im}\,A_{i,k}\left(s,b\right)=|A_{i,k}\left(s,b\right)|^2
+G^{in}_{i,k}(s,b),
\eeq
where $G^{in}_{i,k}$ denotes the contribution of all non 
diffractive inelastic processes,
i.e. it is the summed probability for these final states to be
produced in the scattering of a state $i$ off a state $k$. In \eq{UNIT} 
$\sqrt{s}=W$ is the energy of the colliding hadrons, and $b$ denotes the 
impact  parameter.
A simple solution to \eq{UNIT} at high energies, has the eikonal form 
with an arbitrary opacity $\Omega_{ik}$, where the real 
part of the amplitude is much smaller than the imaginary part.
\beq \label{A}
A_{i,k}(s,b)=i \Lb 1 -\exp\Lb - \Omega_{i,k}(s,b)\Rb\Rb,
\eeq
\beq \label{GIN}
G^{in}_{i,k}(s,b)=1-\exp\Lb - 2\,\Omega_{i,k}(s,b)\Rb.
\eeq
\eq{GIN} implies that 
$P^S_{i,k}=\exp \Lb - 2\,\Omega_{i,k}(s,b) \Rb$
is the probability
 that the initial 
projectiles
$(i,k)$ will reach the final state interaction unchanged, regardless of 
the initial state re-scatterings. 
\par

  %%%%%%%%%%%%%%%%%%%%%%%%%%%%%%%%%%%%%%%%%%%%%%%%%%%%%%%%%%%

 %%%%%%%%%%%%%%%%%%%%%%%%%%%%%%%%%%%%%%%%%%%%%%%%%%%%%%%%%%%%%%%%  

Note, that there is no factor $1/2$. Its absence stems from
 our definition of the dressed Pomeron.

%%%%%%%%%%%%%%%%%%%%%%%%%%%%%%%%%%%%%%%%%%%%%%%%%%%%%%%%%%%%%%%%%
\begin{table}[h]
\begin{tabular}{|l|l|l|l|l|l|l|l|l|}
\hline
\!\!2 ch. model &$\lambda $ & $\phi_0$ ($GeV^{-2}$)&$g_1$
 ($GeV^{-1}$)&$g_2$ ($GeV^{-1}$)& $m(GeV)$ &$m_1(GeV)$&
 $m_2(GeV)$ & $\beta$\\
 \hline
\!  old set & 0.38& 0.0019 & 110.2&  11.2 & 5.25&0.92& 1.9 & 0.58  \\
\hline
\! new  set & 0.325& 0.0021 & 118&  14.7 & 5.45&1.04& 0.47 & 0.52  \\
\hline
\end{tabular}
\caption{Fitted parameters of the model.
 The values of the old set are taken from Ref.\cite{GLM2CH}.
Values of the new set are determined by fitting to data with the 
additional constraint $m_{2}\,\leq$ 1.5 GeV. See Section 4.2.
}
\label{t1}
\end{table}
%%%%%%%%%%%%%%%%%%%%%%%%%%%%%%%%%%%%%%%%%%%%%%%%%%%%%%%%%%%%%%%%%%%%
In the eikonal approximation we replace $ \Omega_{i,k}(s,b)$ by 
\beq \label{EAPR}
 \Omega_{i,k}(s,b)\,\,=\,\int d^2 b'\,d^2 b''\, g_i\Lb \vec{b}'\Rb
 \,G^{\mbox{\tiny dressed}}\Lb T\Lb s, \vec{b}''\Rb\Rb\,g_k\Lb \vec{b} -
 \vec{b}'\ - \vec{b}''\Rb. 
 \eeq
 We propose a more general approach, which takes into account
 new small parameters that result from the fit to the experimental
 data (see Table 1 and \fig{amp}):
 \beq \label{NEWSP}
 G_{3\pom}\Big{/} g_i(b = 0 )\,\ll\,\,1;~~~~~~~~ m\,\gg\, m_1 
~\mbox{and}~m_2.
 \eeq
 
 The second equation in \eq{NEWSP} means that $b''$ in \eq{EAPR}
 is much smaller that $b$ and $ b'$, therefore, \eq{EAPR} can
 be re-written in a simpler form
 \bea \label{EAPR1}
 \Omega_{i,k}(s,b)\,\,&=&\,\Bigg(\int d^2 b''\,G^{\mbox{\tiny
 dressed}}\Lb T\Lb s, \vec{b}''\Rb\Rb\Bigg)\,\int d^2 b' g_i\Lb
 \vec{b}'\Rb \,g_k\Lb \vec{b} - \vec{b}'\Rb \,\nn\\
 &=&\,\tilde{G}^{\mbox{\tiny dressed}}\Lb\bar{T}\Rb\,\,\int d^2
 b' g_i\Lb \vec{b}'\Rb \,g_k\Lb \vec{b} - \vec{b}'\Rb. \eea
 
 Selecting the diagrams using the first equation in \eq{NEWSP}, one
 can see, that the main contribution stems from the net diagrams shown
 in \fig{amp}-b. The sum of these diagrams\cite{GLM2CH} leads to the
 following expression for $ \Omega_{i,k}(s,b)$
 \bea
\Omega\Lb Y; b\Rb~~&=& ~~ \int d^2 b'\,
\,\,\,\frac{ g_i\Lb\vec {b}'\Rb\,g_k\Lb\vec{b} -
 \vec{b}'\Rb\,\tilde{G}^{\mbox{\tiny dressed}}\Lb \bar{T}\Rb
}
{1\,+\,G_{3\pom}\,\tilde{G}^{\mbox{\tiny dressed}}\Lb \bar{T}\Rb\left[
g_i\Lb\vec{b}'\Rb + g_k\Lb\vec{b} - \vec{b}'\Rb\right]} ;\label{OM}\\
g_i\Lb b \Rb~~&=&~~g_i \,S_p\Lb b; m_i \Rb ;\label{g}
\eea
where
$$
S_p\Lb b,m_i\Rb\,=\,\frac{1}{4 \pi} m^3_i \,b \,K_1\Lb m_i b \Rb,
$$
$$
\tilde{G}^{\mbox{\tiny dressed}}\Lb \bar{T}\Rb\,\,=\,\,\int d^2 b
 \,\,G^{\mbox{\tiny dressed}}\Lb T\Lb s, b\Rb\Rb,
$$
where $ T\Lb s, b\Rb$ is given by \eq{T}.

Note  that  $\bar{G}^{\mbox{\tiny dressed}}\Lb \bar T\Rb$ does
 not depend on $b$, and is a function of $
\bar{T} =T\Lb s, b=0 \Rb\,=\,\phi_0 \,e^{0.63\,\lambda Y}$.

In the above formulae the value of the triple BFKL Pomeron vertex
 is known: $G_{3 \pom} = 1.29\,GeV^{-1}$.

\begin{figure}[ht]
    \centering
  \leavevmode
      \includegraphics[width=6cm]{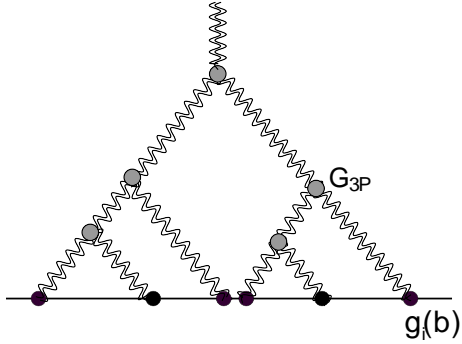}  
      \caption{A typical example of  `fan' diagrams that
 are summed in \protect\eq{NBK}.  }
\label{bk}
   \end{figure}

For further discussion we introduce 

 \beq \label{NBK}
N^{BK}\Lb G^i_\pom\Lb Y,b \Rb\Rb \,\,=\,\,a\,\Lb 1
 - \exp\Lb -  G^i_\pom\Lb Y, b\Rb\Rb\Rb\,\,+\,\,\Lb 1 - a\Rb
\frac{ G^i_\pom\Lb  Y, b\Rb}{1\,+\, G^i_\pom\Lb Y, b\Rb},
\eeq 
 with $a = 0.65$ .
 \eq{NBK} is the analytical approximation for the numerical
 solution to the BK equation\cite{LEPP}. $G_\pom\Lb Y; b\Rb \,=\,\,
 g_i\Lb b \Rb \,\tilde{G}^{\mbox{\tiny dressed}}\Lb \bar{T}\Rb $.
 We recall that the BK equation sums the `fan'  diagrams shown in \fig{bk}.
 
   %%%%%%%%%%%%%%%%%%%%%%%%%%%%%%%%%%%%%%%%%%%%%%%%%%%%%%%%%%%

 %%%%%%%%%%%%%%%%%%%%%%%%%%%%%%%%%%%%%%%%%%%%%%%%%%%%%%%%%%%%%%%%  

   %%%%%%%%%%%%%%%%%%%%%%%%%%%%%%%%%%%%%%%%%%%%%%%%%%%%%%%%%%%%
    
  \section{The main formulae for the survival probability}
  
     %%%%%%%%%%%%%%%%%%%%%%%%%%%%%%%%%%%%%%%%%%%%%%%%%%%%%%%%%%%%  
  
  \subsection{Hard amplitude in the two channel model}
     %%%%%%%%%%%%%%%%%%%%%%%%%%%%%%%%%%%%%%%%%%%%%%%%%%%%%%%%%%%%
     
     The expression for the hard amplitude is known, and it has been
 discussed in great detail (see   Ref.\cite{HLKR}). 
     It has the following general form (see \fig{sp1}-a)
     \beq \label{HAM}
     A^{\mbox{\tiny hard}}\,=\,\pi^2\int d^2 Q_T
 \frac{\bar {\cal M}}{Q^2_T\,\Lb \vec{Q}_T   -
 \vec{p}_{1 T}\Rb^2\,\Lb \vec{Q}_T  + \vec{p}_{2 T}\Rb^2
  } \phi_G\Lb x_1,x'_1, Q^2_T, t_1\Rb \,  \phi_G\Lb x_2,x'_2, Q^2_T, 
t_2\Rb .
     \eeq
     where $Q_T$ is the transverse momentum in the
 gluon loop, $\bar{\cal M}$ is  the color averaged
 amplitude for the process $G G \to X$, where $X$ denotes
 the final state (quark-antiquark jets in \fig{sp1}-a) with mass $M_X$.
   \beq \label{MBAR}
     \bar{\cal M}\,\,=\,\,\frac{2}{M^2_X}\,\frac{1}{N^2_c-1}\sum_{a,b}
 \delta^{a b}\Lb  \vec{Q}^\mu_T   - \vec{p}^\mu_{1 T}\Rb     
     \Lb  \vec{Q}^\nu_T   + \vec{p}^\nu_{2 T}\Rb   \,\Gamma^{a b}_{\mu 
\nu}.    
     \eeq
     $\Gamma^{a b}_{\mu \nu}$ is a vertex for $G G \to X$.
  
  $\phi_G\Lb x_i,x'_i, Q^2_T, t_i\Rb$  denotes the skewed unintegrated
 gluon densities. These functions have been discussed and we refer
the reader
 to Ref.\cite{HLKR}. The $t_i$ dependence,
  is of great importance for the calculation of the
 survival probability \cite{GLMSP1,GLMSP2} .
 We  show below that the essential $t_i$  turns
 out to be small in our estimates, and therefore, we have to rely on some 
input  from
 non-perturbative QCD. Our assumption is that at small $t_i$ we can 
factorize
 the unitegrated gluon density as
  \beq \label{FA}
  \phi_G\Lb x_i,x'_i, Q^2_T, t_i\Rb\,\,=\,\,\tilde{\phi}_G\Lb x_i,x'_i,
 Q^2_T\Rb\,\Gamma\Lb t_i\Rb\,\,~~\xrightarrow{\mbox{ impact parameter
 image}}\,\,~~\tilde{\phi}_G\Lb x_i,x'_i, Q^2_T\Rb\,S^h\Lb b \Rb.
  \eeq 
  
   Making  the assumption,   the hard amplitude at fixed impact 
parameter 
$b$, 
has the form 
     \beq \label{HAM1}     
      A^{\mbox{\tiny hard}}\,=\,     A^{\mbox{\tiny hard}}\Lb s;
 b,p_1,p_2\Rb \,\int d^2 b' \,S^h\Lb b'\Rb\,S^h\Lb \vec{b} - \vec{b}'\Rb.
     \eeq

\begin{figure}[ht]
    \centering
  \leavevmode
      \includegraphics[width=10cm]{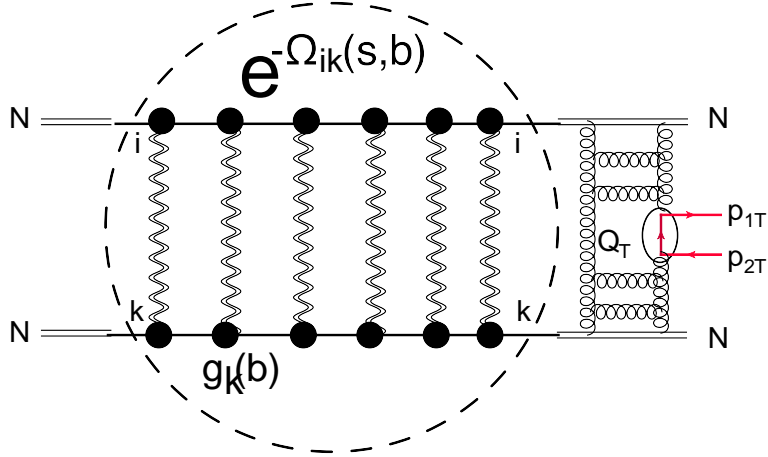}  
      \caption{Survival probability in two channel model.}
\label{sph}
   \end{figure}

 The advantage of our technique is that it is based on the
 CGC/saturation approach, and the unintegrated  structure
 functions $\phi_G\Lb x_i,x'_i, Q^2_T, t_i\Rb$,  can be calculated in
 this framework. In the two channel model we have two unintegrated 
structure
 functions (see \fig{sph}): 

 \bea \label{HARDAM}
 \phi_{1\to \mbox{proton}} &\,\propto \,\,&\alpha \,g_{1}\Lb b\Rb\,\equiv\,
 S^h_1\Lb b\Rb;\nn\\
  \phi_{2\to \mbox{proton}}&\,\propto \,\,&\beta \,g_{2}\Lb b\Rb\,\equiv\,
 S^h_2\Lb b\Rb;\nn
  \eea

  We extract the $b$ dependence of hard amplitudes  i,  from the
 experimental data for diffractive production of vector mesons in
 deep inelastic scattering (DIS). Presenting the $t$-dependance of
 the measure differential cross section in the form
 \beq \label{B}
\frac{d \sigma\Lb \gamma^* + p \,\to\,V  + p \Rb}{d t}\Bigg{/}\frac{d
 \sigma\Lb \gamma^* + p \,\to\,V  + p \Rb}{d t}\big{|}_{t=0}
\,\,=\,\,e^{ - B^h\,|t|}.
 \eeq 
 In QCD 
 \bea \label{NIMVP}
  \frac{d \sigma\Lb \gamma^* + p \,\to\,V  + p \Rb}{d t}
 \,&\propto& \,\phi^2_G\Lb x_{Bj},x'_{Bj}, Q^2, t\Rb  \nn\\
   &\propto&\Lb  \int e^{i \vec{Q}_T \cdot \vec{b}}\Lb\alpha^2 
 g_1 S_p\Lb b, m_1\Rb\,+\,\beta^2 \,g_2\,S_p\Lb b, m_2\Rb\Rb\Rb^2,
   \eea
 where $t = - Q^2_T$.    
 
\begin{figure}[ht]
    \centering
  \leavevmode
      \includegraphics[width=9cm]{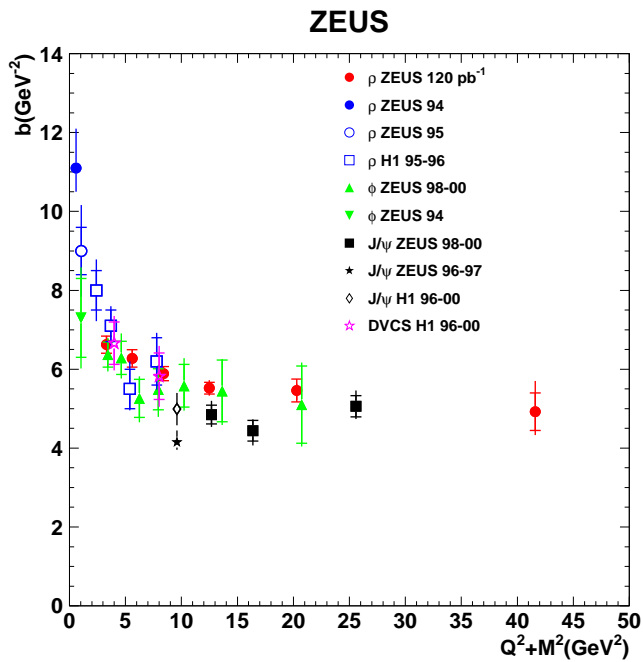}  
      \caption{Compilation of experimental data on the slope 
of the diffractively produced produced vector mesons at HERA. The figure
 is taken from Ref.\cite{SLOPEZEUS}
   }
\label{slope}
   \end{figure}

 The value of the slope $B^h$ can be calculated and it is equal to
 \beq \label{BHARD}
 B^h\,=\,\h\,\frac{\int b^2\, d^2 b \Lb\alpha^2 
 g_1 S_p\Lb b, m_1\Rb\,+\,\beta^2 \,g_2\,S_p\Lb b,
 m_2\Rb\Rb}{\int d^2 b \Lb\alpha^2  g_1 S_p\Lb b,
 m_1\Rb\,+\,\beta^2 \,g_2\,S_p\Lb b, m_2\Rb\Rb }.
 \eeq
 Using the  parameters of Table 1, we find  $B^h \approx 4.5\,GeV^{-2}$.
 From \fig{slope} one can see that $B^h \,\to \,4 \div 5 \,GeV^{-2}$.
 Therefore, the $b$ dependence obtained from our approach, is in 
 accord with the HERA experimental data.

  %%%%%%%%%%%%%%%%%%%%%%%%%%%%%%%%%%%%%%%%%%%%%%%%%%%%%%%%%%%

 %%%%%%%%%%%%%%%%%%%%%%%%%%%%%%%%%%%%%%%%%%%%%%%%%%%%%%%%%%%%%%%%

     Generally speaking, both $B^h$'s depend on energy. Indeed, in
 Regge theory \cite{COL}, the scattering amplitude 
     $A^{\mbox{\tiny hard}}\,\,\propto\,\,s^{\alpha_\pom(t)}$
 with $\alpha_\pom\Lb t \Rb\,=\,\,\alpha_\pom\Lb 0\Rb\,\,+\,\,\alpha'_\pom
 \,\ln\Lb s/s_0\Rb\,t\,=\,1\,+\,\Delta\,+ \,\,+\,\,\alpha'_\pom
 \,\ln\Lb s/s_0\Rb\,t\,$. For  hard processes we do not expect
  Pomeron trajectories
  with $\alpha'_\pom \neq 0$. However,  the effective $\alpha'_\pom $
 is due to shadowing corrections. The hard amplitude has
 the following generic form
  \beq \label{HA25}
   A^{\mbox{\tiny hard}}\,\,\propto\,s^\Delta \,e^{- \frac{b^2}{2 B^h}}.    
   \eeq
   At large $b$ this amplitude is small. At some value of $b=b_0(s)$,
 $  A^{\mbox{\tiny hard}}\Lb s, b\Rb \sim 1$. This equation leads to
   \beq \label{HA26}
   e^{- \frac{b_0^2}{2 B^h}}    = f \,\leq 1; ~~~~~~ b^2_0(s)\,=\,2\,B^h
 \,\Delta \ln\Lb s/s_0\Rb.
   \eeq
   
    Due to unitarity (see \eq{UNIT}) the amplitude cannot exceed unity. 
 Therefore, at $b \,\leq\,b_0(s)$,   $A^{\mbox{\tiny hard}}\Lb s, b\Rb    
  \,\propto \Theta\Lb b_0(s) - b\Rb$ where $\Theta(z)$ is a step function.
 On the other hand, the $t$ slope of the amplitude, is equal to  $B = 4 
\langle
 b^2 \rangle = 8 b^2_0(s)$. The last equation stems from $ A^{\mbox{\tiny
 hard}}\Lb s, b\Rb      \,\propto \Theta\Lb b_0(s) - b\Rb$. Finally, the
 t-slope for the scattering amplitude is proportional to $\ln(s/s_0)$, viz.
 $B\,=\,\frac{1}{4}\,B^h \,\Delta \ln\Lb s/s_0\Rb   $ or
   \beq \label{HA27}
   \alpha^{'\,eff}_\pom\,\,=\,\,\frac{1}{4} \Delta \,B^h_{el,0}\, ,
   \eeq
   where $B^h_0$ is the slope for the cross section at $s=s_0$.
 Choosing $s_0 = 1 \,GeV^2$ and $\Delta = 0.2$\footnote{ Such
 values of $\Delta$ come both from  experiment \cite{SLOPEZEUS,
 SLOPEH1} and from  theoretical estimates.}   we obtain $ B^h_0
 \approx 3.2 \,GeV^{-2}$
   and   $   \alpha_\pom^{'\,eff} \,=\,0.154 \,GeV^{-2}$. While 
 the HERA experiment gives \cite{SLOPEZEUS,SLOPEH1}
   $ B^h_{el}\,\,=\,\,4.6 \pm 0.06\,\,+\,\,4\,\Lb 0.164 \pm 0.41\Rb\ln\Lb
 W/W_0\Rb$. We believe that our approach provides a reasonable estimate 
of, 
and  an appropriate method to
 understand the energy behavior of the hard amplitude. Note,
 that $B^h_0  \approx 3.2 \,GeV^{-2}$ comes from the experimental
 formulae changing $W_0 = 90 \,GeV$ to $W_0= 1\,GeV$.   
              
   %%%%%%%%%%%%%%%%%%%%%%%%%%%%%%%%%%%%%%%%%%%%%%%%%%%%%%%%%%%

 %%%%%%%%%%%%%%%%%%%%%%%%%%%%%%%%%%%%%%%%%%%%%% %%%%%%%%%%%%%%%%%%  
 
 Bearing in mind this estimate, we find the following hierarchy
 of  transverse distances, in our approach for high energy
 scattering (see Table 1)
 \beq \label{HIERB}
 \frac{4}{m_1^2}\,>\, \, \,B^{\mbox{\tiny hard}
 }\,\,\gg\,\frac{1}{m^2},
 \eeq
 where $4/m^2_i$ is the typical slope for  $g_i\Lb b \Rb$.
 
     %%%%%%%%%%%%%%%%%%%%%%%%%%%%%%%%%%%%%%%%%%%%%%%%%%%%%%%%%%%%  
  
  \subsection{Survival probability: eikonal approach}
     %%%%%%%%%%%%%%%%%%%%%%%%%%%%%%%%%%%%%%%%%%%%%%%%%%%%%%%%%%%%
  Central  exclusive production (CEP) has the
  typical form
\beq \label{CEP}
p\,~~+~~\,p\,\,\to\,\,p \Lb q_{1,T}\Rb\,~~+~~\Big[LRG \Big]
\,~~~+~~X\Lb M_X\Rb~~+~~\,\,\Big[LRG\Big]\,\,~~~+~~p \Lb q_{2,T}\Rb,\eeq
where LRG denotes the large rapidity gap in which  no hadrons
 are produced. We cannot restrict ourselves to the hard amplitude 
(see
 \fig{sp1}-a) to describe this reaction. Indeed, \eq{HAM} gives the
 amplitude of CEP, but only for one parton shower. However, the production
 of many parton showers , shown in \fig{sp1}-c, will contaminate  the 
LRG's,
 and these have to be eliminated, in order  to obtain the correct cross 
section for the
 reaction of \eq{CEP}. In  simple eikonal models, such a suppression
 stems from the diagrams shown in \fig{sp1}-b. 

For this simple case, we can derive the formula for the survival
 probability, using two different approaches. 
The first one,  relies on the s-channel unitarity constraint (see 
\eq{UNIT}).
  In the eikonal approach the contribution of all inelastic states is 
given
 by \eq{GIN}
\beq \label{GIN1}
G_{in}\Lb s, b\Rb\,\,=\,\,1\,\,-\,\,\exp\Lb - \,2\,\Omega\Lb s, b\Rb\Rb.
\eeq
From \eq{GIN1} we see that multiplying the hard cross section by the
 factor $\exp\Lb - \,2\,\Omega\Lb s, b\Rb\Rb$,
we obtain the probability that the process has no inelastic
 production in the entire kinematic rapidity region \cite{BJ,DKS,GLMSP1}.
 Therefore, the survival probability factor $\langle S^2\rangle$
 takes the form
\bea \label{EIKSP}
\langle S^2\rangle\,\,&=&\,\,\frac{\int d^2 b \, e^{ - 2\,\Omega\Lb s,
 b\Rb}\,\Big{|}A^{\mbox{\tiny hard}}\Lb s; b,p_1,p_2\Rb \,\int d^2 b'
 \,S_h\Lb b'\Rb\,S_h\Lb \vec{b} - \vec{b}'\Rb\Big{|}^2 }{ \int d^2 b\,\Big{|}
 \int d^2 b' \,A^{\mbox{\tiny hard}}\Lb s; b,p_1,p_2\Rb\,S_h\Lb b'\Rb\,S_h\Lb
 \vec{b} - \vec{b}'\Rb\Big{|}^2}\nn\\
& =&\,\,\frac{\int d^2 b \, e^{ - 2\,\Omega\Lb s, b\Rb}\,\Big{|} \,\int d^2 b'
 \,S_h\Lb b'\Rb\,S_h\Lb \vec{b} - \vec{b}'\Rb\Big{|} }{ \int d^2 b\,\Big{|} 
 \,\int d^2 b' \,S_h\Lb b'\Rb\,S_h\Lb \vec{b} - \vec{b}'\Rb\Big{|}^2}.
\eea
The second derivation  is based on summing the Pomeron diagrams 
 of \fig{sp1}-b,   introducing  $\Omega = g\Lb 
b\Rb^2 
\tilde{G}^{\mbox{\tiny
 dressed}}\Lb s\Rb$ in \eq{GIN1}. The eikonal amplitude can be
 written as (see \eq{A})
\beq \label{A1}
i \Big( 1 \,\,\,-\,\,\exp\Lb - \Omega\Lb s, b\Rb\Rb\Big)\,\,=\,
\,i\,\sum^\infty_{n = 1}\,\Lb - 1 \Rb^{n - 1}\,\frac{\Omega^n\Lb s, 
b\Rb}{n!}.
\eeq
 In each term with the exchange of $n$-Pomerons, we need to replace one of
 these Pomerons by the hard amplitude.
Such a replacement leads   to the following sum
\beq \label{A2}
i\,\sum^\infty_{n = 1}\,\Lb - 1 \Rb^{n - 1}\,\frac{n\,\,\Omega^n\Lb s,
 b\Rb}{n!}\,A^{\mbox{\tiny hard}}\,\,=\,\,i\,e^{-\Omega\Lb s,
 b\Rb}A^{\mbox{\tiny hard}}.
\eeq
Multiplying this amplitude by its complex conjugate,
 and integrating over $b$ we obtain \eq{EIKSP}.

      %%%%%%%%%%%%%%%%%%%%%%%%%%%%%%%%%%%%%%%%%%%%%%%%%%%%%%%%%%%%  
  
  \subsection{Survival probability: enhanced diagrams}
\begin{figure}[ht]
    \centering
  \leavevmode
      \includegraphics[width=14cm]{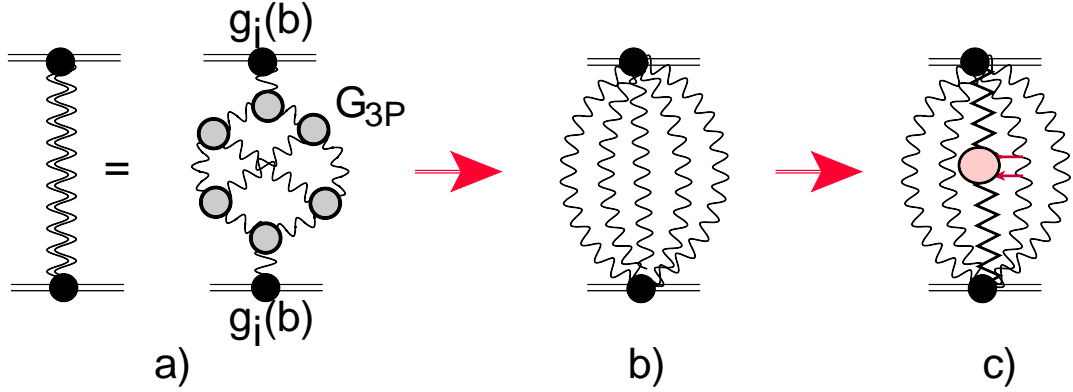}  
      \caption{\fig{enh}-a shows the set of the diagrams for the BFKL
 Pomeron calculus that lead to the resulting  Pomeron Green function,
  after integrating
 over rapidities of the triple Pomeron vertices, 
 can be re-written in the form of \fig{enh}-b.  \fig{enh}-c 
illustrates 
   that to calculate the survival probability,
we need to replace  one of  BFKL Pomerons by the hard amplitude. 
}
\label{enh}
   \end{figure}

     %%%%%%%%%%%%%%%%%%%%%%%%%%%%%%%%%%%%%%%%%%%%%%%%%%%%%%%%%%%%
   At first sight \eq{EIKSP}, provides the answer for the case of  
eikonal rescattering. However, this is not correct, since the dressed
 BFKL Pomeron Green function  is the sum of   enhanced diagrams of 
\fig{amp}-a
 and \fig{enh}-a. To find the survival probability, we need to replace
 one of the Pomeron lines in \fig{enh}-a  by the hard amplitude.   As
 was noticed in Ref.\cite{LMP} the enhanced diagrams can be reduced to
 a sum of diagrams which have a general form
   \beq \label{ENH1}
 G^{\mbox{\tiny dressed}}\Lb T\Rb\,\,=\,\,\sum_{n=1}^\infty\,(-1)^{n-1}\,
\Gamma^2\Lb P \to n P\Rb   \,T^n ,
   \eeq  
    after integration over positions in rapidity, of the triple Pomeron 
vertices.
  The vertices $ \Gamma\Lb P \to n P\Rb $
   can  easily be found from \eq{G}. To obtain the survival probability, 
we 
need to replace 
 $T$ in \eq{ENH1} by the hard amplitude : viz
   \bea \label{ENH2}
 &&G^{\mbox{\tiny hard}}\Lb Y, b\Rb \,\equiv\,  
 A^{\mbox{\tiny hard}}_\pom\sum^\infty_{n=1}\Lb -1 \Rb^n \,n\,
\Gamma^2\Lb P \to n P\Rb   \,T^{n - 1}\, \nn\\
&& \to\, A^{\mbox{\tiny hard}}_\pom\Bigg\{ a^2\,e^{-T}\,-\,(1 -
 a)\Big( \frac{1 - a}{T^2} - \frac{2 a}{\Lb 1 + T\Rb^2}\Big)+
 (1 - a)^2 \frac{1 + T}{T^3} e^{\frac{1}{T}}\Gamma_0\Lb
 \frac{1}{T}\Rb\Bigg\}.    \eea

It should be noted that $A^{\mbox{\tiny hard}}_\pom$ is not the same as
 in \eq{EIKSP}, and its $b$ distribution has a typical value of $b \propto
 1/m$. In other words $A^{\mbox{\tiny hard}}_\pom\,\propto\,S\Lb b\Rb $. 
  \eq{ENH2} leads to the following  contribution  
 \bea
 \Omega^{ \mbox{\tiny hard}}_{i k}&=&\,\int\!\! d^2 b'\,d^2 b''\,g_i\Lb b'\Rb 
  \,G^{\mbox{\tiny hard}}\Lb Y, b''\Rb   \, g_k\Lb \vec{b} - \vec{b}'  - 
\vec{b}''\Rb \to \int \!\!d^2 b'\,g_i\Lb b'\Rb\,g_k\Lb \vec{b} - \vec{b}'
\Rb\tilde{G}^{\mbox{\tiny hard}}\Lb Y\Rb, \label{ENH3} \\
\tilde{G}^{\mbox{\tiny hard}}\Lb Y\Rb  &=&\int\!\! d^2 b''
 \,S\Lb b''\Rb\Bigg\{ a^2\,e^{-T}-(1 - a)\Big( \frac{1 - a}{T^2}
 -\frac{2 a}{\Lb 1 + T\Rb^2}\Big)\,+\,(1 - a)^2 \frac{1 + T}{T^3}
 e^{\frac{1}{T}}\Gamma_0\Lb \frac{1}{T}\Rb\Bigg\}.  \label{ENH4} \eea
      
        %%%%%%%%%%%%%%%%%%%%%%%%%%%%%%%%%%%%%%%%%%%%%%%%%%%%%%%%%%%

 %%%%%%%%%%%%%%%%%%%%%%%%%%%%%%%%%%%%%%%%%%%%%% %%%%%%%%%%%%%%%%%%   
   Inspecting \eq{ENH3} we note that at $T \to 0$,  \eq{ENH3} leads to   
 $\Omega^{ \mbox{\tiny hard}}_{i k}\,\,\to \,\int\!\! d^2 b'\,d^2 b''\,g_i
\Lb b'\,\Rb\,g_k\Lb \vec{b} - \vec{b}'\Rb$, which coincides with our hard
 amplitude introduced in \eq{HARDAM}. Using the notation  in this 
equation,
 the expression for  $   \Omega^{ \mbox{\tiny hard}}_{i k}$ takes   the 
final
 form:
   \beq \label{ENH5}
    \Omega^{ \mbox{\tiny hard}}_{i  k} \,\,=\,\,\,\int d^2 b'\,S_i^h\Lb
 b'\Rb\,S_k^h\Lb \vec{b} - \vec{b}'\Rb\tilde{G}^{\mbox{\tiny hard}}\Lb 
Y\Rb.   
    \eeq
   
      %%%%%%%%%%%%%%%%%%%%%%%%%%%%%%%%%%%%%%%%%%%%%%%
  
  \subsection{Survival probability: general formulae}
     %%%%%%%%%%%%%%%%%%%%%%%%%%%%%%%%%%%%%%%%%%%%%%%  
 
      %%%%%%%%%%%%%%%%%%%%%%%%%%%%%%%%%%%%%%%%%%%%%%%  
  
  \subsubsection{Survival probability:  eikonal formula for two channel model}
     %%%%%%%%%%%%%%%%%%%%%%%%%%%%%%%%%%%%%%%%%%%%%%%%%%%%%%%%%%        
         The structure of the formula for the survival probability
 is shown in \fig{sph}. The amplitude for the reaction of \eq{CEP}
         can be written in the form
         \bea \label{EIKSP1}
  &&       A^{full}\Lb s, q_{1,T},q_{2,T}, p_1,p_1\Rb\,\,=\,\,\int d^2 b'
 d^2 b \,e^{i \vec{q}_{1,T}\cdot \vec{b'}} \,e^{i \vec{q}_{2,T}\cdot\Lb
 \vec{b} -  \vec{b'}\Rb}  \,       
          \tilde{G}^{\mbox{\tiny hard}}\Lb Y\Rb \nn\\
          &&\times\Bigg\{ \alpha^2 e^{ - \Omega_{11}\Lb, s,b\Rb}
 S^h_1\Lb b'\Rb \,S_1^h\Lb \vec{b} - \vec{b}'\Rb  \,+\,\beta^2 
 e^{- \Omega_{22}\Lb, s,b\Rb}S^h_2\Lb b'\Rb \,S_2^h\Lb \vec{b}
 - \vec{b}'\Rb\nn\\
         &&  \,\,+\,\,\alpha\, \beta\Lb e^{ - \Omega_{12}\Lb,
 s,b\Rb}S^h_1\Lb b'\Rb \,S_2^h\Lb \vec{b} - \vec{b}'\Rb   \,+\,
  e^{ - \Omega_{21}\Lb, s,b\Rb}S^h_2\Lb b'\Rb \,S_1^h\Lb \vec{b}
 - \vec{b}'\Rb\Rb  \Bigg\} .       \eea
         
     The survival probability for the cross section $d^2 
\sigma/(d t_1\,d t_2   )$ is equal
     \beq \label{EIKSP2} 
         \langle S^2\rangle =   \Big{|}  A^{full}\Lb s,
 q_{1,T},q_{2,T}, p_1,p_1\Rb   \Big{|}^2\bigg{/} \Big{|}
  A^{hard}\Lb s, q_{1,T},q_{2,T}, p_1,p_1\Rb   \Big{|}^2,
         \eeq
         where 
         \bea \label{EIKSP3}
  &&       A^{hard}\Lb s, q_{1,T},q_{2,T}, p_1,p_1\Rb\,\,=\,\,\int d^2 b'
 d^2 b \,e^{i \vec{q}_{1,T}\cdot \vec{b'}} \,e^{i \vec{q}_{2,T}\cdot\Lb
 \vec{b} -  \vec{b'}\Rb}  \,       
           \\
          &&\times\Bigg\{ \alpha^2 S^h_1\Lb b'\Rb \,S_1^h\Lb
 \vec{b} - \vec{b}'\Rb  \,+\,\beta^2  S^h_2\Lb b'\Rb \,S_2^h\Lb \vec{b}
 - \vec{b}'\Rb\,\,+\,\,\alpha\, \beta \Lb S^h_1\Lb b'\Rb \,S_2^h\Lb 
\vec{b}
 - \vec{b}'\Rb   \,+\, S^h_2\Lb b'\Rb \,S_1^h\Lb \vec{b} - \vec{b}'\Rb\Rb 
 \Bigg\} .  \nn        \eea
          
    However, if we are interested in  cross sections that are
 integrated over $d^2 q_{1,T}$ and $d^2 q_{2,T}$, the expression
 for $\langle S^2\rangle$ can be simplified, and it has the form
    
   \beq \label{EIKSP4}
   \langle S^2\rangle = N\Lb s, M_x,p_1,p_2\Rb\Big{/} 
  D\Lb S, M_x,p_1,p_2\Rb,     
   \eeq
   with
    \bea \label{EIKSP5}
  &&   N\Lb s, M_x,p_1,p_2\Rb  \,\,=\,\,\int  d^2 b \        
         \Lb  \tilde{G}^{\mbox{\tiny hard}}\Lb Y\Rb \Rb^2\nn\\
          &&\times\Bigg\{ \int d^2 b'\,\Bigg( \alpha^2 e^{ -
 \Omega_{11}\Lb, s,b\Rb} S^h_1\Lb b'\Rb \,S_1^h\Lb \vec{b} -
 \vec{b}'\Rb  \,+\,\beta^2  e^{- \Omega_{22}\Lb, s,b\Rb}S^h_2\Lb b'\Rb
 \,S_2^h\Lb \vec{b} - \vec{b}'\Rb\nn\\
         &&  \,\,+\,\,\alpha\, \beta\Lb e^{ - \Omega_{12}\Lb,
 s,b\Rb}S^h_1\Lb b'\Rb \,S_2^h\Lb \vec{b} - \vec{b}'\Rb   \,+\,
  e^{ - \Omega_{21}\Lb, s,b\Rb}S^h_2\Lb b'\Rb \,S_1^h\Lb \vec{b}
 - \vec{b}'\Rb\Rb \Bigg) \Bigg\}^2,
         \eea

                      and
                      
    \bea \label{EIKSP6}   
    D\Lb s, M_x,p_1,p_2\Rb  \,\,&=&\,\,\int  d^2 b \,\Bigg\{\int d^2 b'\,
\Bigg( \alpha^2 S^h_1\Lb b'\Rb \,S_1^h\Lb \vec{b} - \vec{b}'\Rb  \,+\,\beta^2
  S^h_2\Lb b'\Rb \,S_2^h\Lb \vec{b} - \vec{b}'\Rb\nn\\
         & +&\,\,\alpha\, \beta\Lb  S^h_1\Lb b'\Rb \,S_2^h\Lb \vec{b} -
 \vec{b}'\Rb   \,+\, S^h_2\Lb b'\Rb \,S_1^h\Lb \vec{b} - \vec{b}'\Rb\Rb
\Bigg) \Bigg\}^2.
         \eea

      %%%%%%%%%%%%%%%%%%%%%%%%%%%%%%%%%%%%%%%%%%%%%%%%%%%%%%%%%%%%  
  \begin{boldmath}
  \subsubsection{ General  case: $ \Omega^{\mbox{\tiny hard}}$}
  \end{boldmath}
     %%%%%%%%%%%%%%%%%%%%%%%%%%%%%%%%%%%%%%%%%%%%%%%%%%%%%%%%%%  
       
       The first problem that we need to solve, is to find a more general
 expression for $\Omega^{\mbox{\tiny hard}}$, than we  have obtained in 
\eq{ENH5}. 
 \eq{OM} sums net diagrams, and they can be re-written in the same form
 as the enhanced ones\cite{LMP}. \eq{ENH1} is replaced by
       \beq \label{OMG1}
        \Omega_{i k}\,\,=\,\,   \sum^\infty_{n=1}\,\Lb - 1 \Rb^{n - 1}
 \Gamma\Lb i \to n \pom\Rb  \, \Gamma\Lb k \to n \pom\Rb \Lb
 \tilde{G}^{\mbox{\tiny dressed}}\Rb^n .
        \eeq
        From \eq{OMG1} we obtain
      \beq \label{OMG2}      
      \Omega^{\mbox{\tiny hard}}_{i k}\,=\,A^{\mbox{\tiny hard}}
\,\sum^\infty_{n = 0}   \Gamma\Lb i \to (n  + 1) \pom\Rb  \,
 \Gamma\Lb k \to (n  + 1)\pom\Rb \Lb \tilde{G}^{\mbox{\tiny dressed}}\Rb^n 
.  
      \eeq
     Using \eq{OM} and \eq{OMG2} we obtain
       
       \beq               \label{OMG3}
       \Omega^{\mbox{\tiny hard}}_{i k}\Lb Y; b\Rb =  \int d^2 b'\,
\,\,\,\frac{ S^i_h\Lb\vec {b}'\Rb\,S^k_h\Lb\vec{b} - \vec{b}'\Rb}
{\Bigg(1\,+\,G_{3\pom}\,\tilde{G}^{\mbox{\tiny dressed}}\Lb T\Rb\left[
g_i\Lb\vec{b}'\Rb + g_k\Lb\vec{b} - \vec{b}'\Rb\right]\Bigg)^2}.
\eeq                   
        %%%%%%%%%%%%%%%%%%%%%%%%%%%%%%%%%%%%%%%%% %%%%%%%%%%%%%%%%%%%      
     \begin{figure}[ht]
    \centering
  \leavevmode
      \includegraphics[width=14cm]{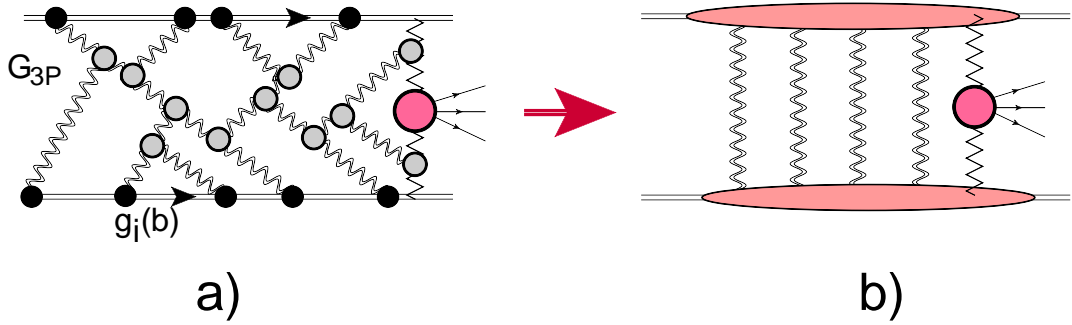}  
      \caption{\fig{sp}-a shows the set of the net 
 diagrams in the BFKL Pomeron calculus, that lead to
 the resulting  survival probability  in the framework of
 high energy QCD. After integration over positions of $G_{3 \pom}$ in 
rapidity,
  the sum of the diagrams reduces to \fig{sp}-b. }
\label{sp}
   \end{figure}

 %%%%%%%%%%%%%%%%%%%%%%%%%%%%%%%%%%%%%%%%%%%%%%   %%%%%%%%%%%%%%%%%%%      

 Taking into account the enhanced diagrams for $\tilde{G}^{\mbox{\tiny
 dressed}}$ (see \eq{ENH3}) we obtain the final form for 
    $ \Omega^{\mbox{\tiny hard}} _{i k}$:

       \beq              \label{OMG4}
       \Omega^{\mbox{\tiny hard}}_{i k}\Lb Y; b\Rb = \int d^2 b'
\,\tilde{G}^{\mbox{\tiny hard}}\Lb Y\Rb\,\,\,\frac{ S^i_h\Lb\vec
 {b}'\Rb\,S^k_h\Lb\vec{b} - \vec{b}'\Rb}
{\Bigg(1\,+\,G_{3\pom}\,\tilde{G}^{\mbox{\tiny dressed}}\Lb T\Rb\left[
g_i\Lb\vec{b}'\Rb + g_k\Lb\vec{b} - \vec{b}'\Rb\right]\Bigg)^2}
  =  \tilde{G}^{\mbox{\tiny hard}}\Lb Y\Rb\, \bar{ \Omega}^{\mbox{\tiny
 hard}}_{i k}.
\eeq    
      %%%%%%%%%%%%%%%%%%%%%%%%%%%%%%%%%%%%%%%%%%%%%%%%%%%%%%%%%%%%  

  \subsubsection{Final formula}

     %%%%%%%%%%%%%%%%%%%%%%%%%%%%%%%%%%%%%%%%%%%%%%%%%%%%%%%%%%  
   Finally, to obtain the general formula for the survival probablity we
 need in \eq{EIKSP5}, we replace \\
  $ \int d^2 b'\, S^i_h\Lb\vec {b}'\Rb\,S^k_h\Lb\vec{b} - \vec{b}'\Rb$ by
 $   \bar{ \Omega}^{\mbox{\tiny hard}}_{i k}\Lb Y, b\Rb$.
   Therefore, the survival probability is equal to
      \beq \label{FF1}
   \langle S^2\rangle = N\Lb s, M_x,p_1,p_2\Rb\Big{/} 
  D\Lb s, M_x,p_1,p_2\Rb,     
   \eeq
   with
    \bea \label{FF2}
    N\Lb s, M_x,p_1,p_2\Rb  \,\,&=& \,\,\int d^2 b        
  \Lb  \tilde{G}^{\mbox{\tiny hard}}\Lb Y\Rb \Rb^2 \times\Bigg\{
 \alpha^2 e^{ - \Omega_{11}\Lb s,b\Rb}   \bar{ \Omega}^{\mbox{
\tiny hard}}_{1 1}\Lb Y, b\Rb   \,+\,\beta^2  e^{- \Omega_{22}\Lb
 s,b\Rb}  \bar{ \Omega}^{\mbox{\tiny hard}}_{2 2}\Lb Y, b\Rb \nn\\
&+&\,\,\alpha\, \beta\Lb e^{ - \Omega_{12}\Lb s,b\Rb}  \bar{
 \Omega}^{\mbox{\tiny hard}}_{1 2 }\Lb Y, b\Rb\,+\,  e^{ -
 \Omega_{21}\Lb s,b\Rb}  \bar{ \Omega}^{\mbox{\tiny hard}}_{2
 1 }\Lb Y, b\Rb\Rb\Bigg\}^2.
         \eea   
     
     while $ D\Lb s, M_x,p_1,p_2\Rb$ remains the same  as in \eq{EIKSP6}.
     
     From \eq{FF1} and \eq{FF2} we note that $\langle S^2\rangle
 \,\propto\,  \Lb  \tilde{G}^{\mbox{\tiny hard}}\Lb Y\Rb \Rb^2$.
     This factor  takes into account the contribution from 
the enhanced diagrams. \fig{spenfig} shows that on its own, it leads to 
a smaller survival probability .
       %%%%%%%%%%%%%%%%%%%%%%%%%%%%%%%%%%%%%%%%%%%%     
   \section{Numerical estimates}
      %%%%%%%%%%%%%%%%%%%%%%%%%%%%%%%%%%%%%%%%%%%%
      
      \subsection{Survival probability in our model}
   %%%%%%%%%%%%%%%%%%%%%%%%%%%%%%%%%%%%%%%%%%%%
        
        Our estimates for the survival probability are shown in \fig{fin}.

     \begin{figure}[ht]
    \centering
  \leavevmode
      \includegraphics[width=10cm]{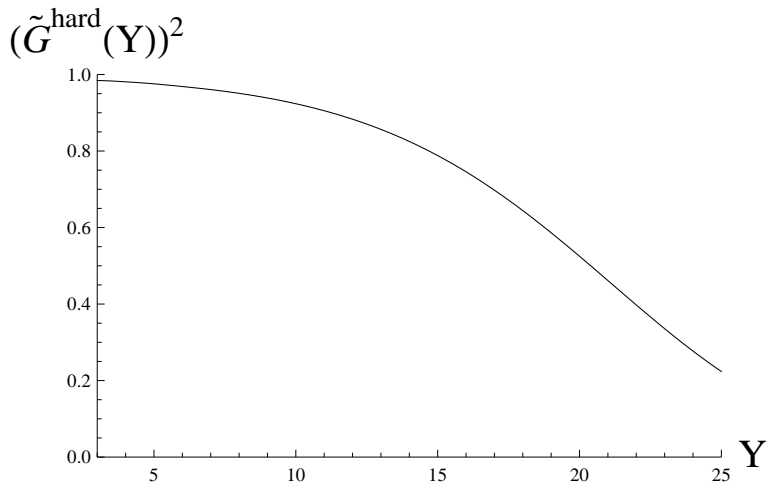}  
      \caption{ The suppression factor $\Lb 
 \tilde{G}^{\mbox{\tiny hard}}\Lb Y\Rb \Rb^2$ which
 includes the contribution of the enhanced diagrams. }
\label{spenfig}
   \end{figure}

 %%%%%%%%%%%%%%%%%%%%%%%%%%%%%%%%%%%%%%%%%%%%%% %%
              %%%%%%%%%%%%%%%%%%%%%%%%%%%%%%%%%%%%%%%%%%%%
     \begin{figure}[ht]
    \centering
  \leavevmode
      \includegraphics[width=14cm]{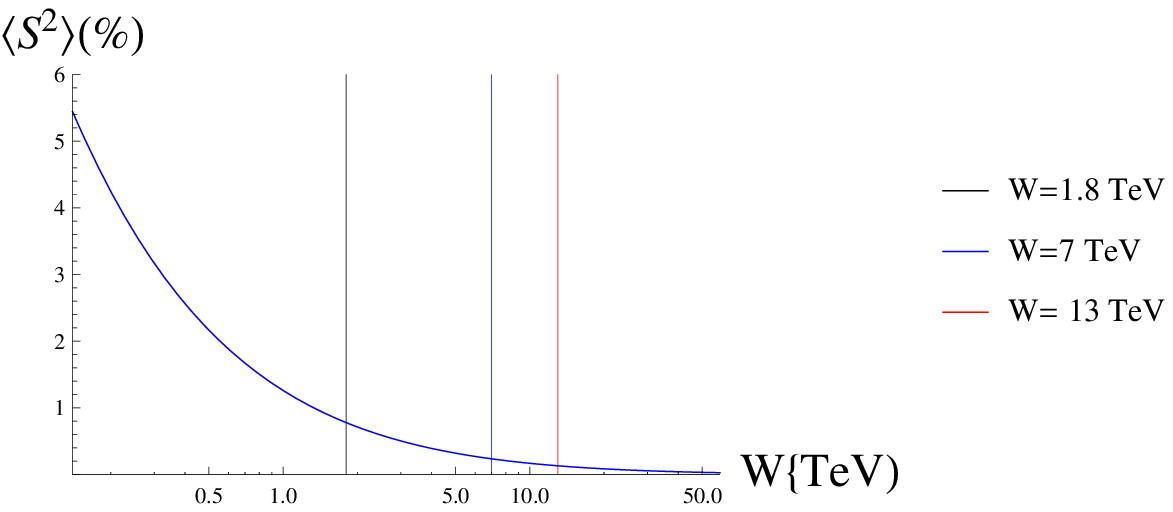}  
      \caption{ $\langle S^2\rangle$ of \protect\eq{FF1} versus $W$. }
\label{fin}
   \end{figure}

 %%%%%%%%%%%%%%%%%%%%%%%%%%%%%%%%%%%%%%%%%%%%%% %% 
  We predict rather small values for the
 survival probability. Such small values have been discussed
 peviously (see Ref.\cite{GLMSP2}), however, in the present model we have 
 a different source  for this small number. In Ref.\cite{GLMSP2} $
 \langle S^2\rangle$ turns out to be small, due to contribution
 of the enhanced diagram, while in the present  model the enhanced 
diagrams
 give a suppression factor which is moderate (see \fig{spenfig}).
 The main cause for the small value of  $\langle S^2\rangle$,
 is  the $b$ dependence of the hard amplitude.
 As we have mentioned  our $b$ dependence stems from the
 description of the soft high energy data, based on CGC/saturation
 approach, in which we do not  introduce a special soft amplitude.
 In our approach we are only dealing  with hard (semi-hard) amplitudes,
 which
 provide a smooth matching of the `soft' interaction with the  'hard' 
one. 
     %%%%%%%%%%%%%%%%%%%%%%%%%%%%%%%%%%%%%%%%%%%%
     \subsection{Importance of $b$-dependence of the hard amplitude.}
    %%%%%%%%%%%%%%%%%%%%%%%%%%%%%%%%%%%%%%%%%%%%
    
\begin{figure}[ht]
    \centering
    \begin{tabular}{c}
  \leavevmode
      \includegraphics[width=14cm]{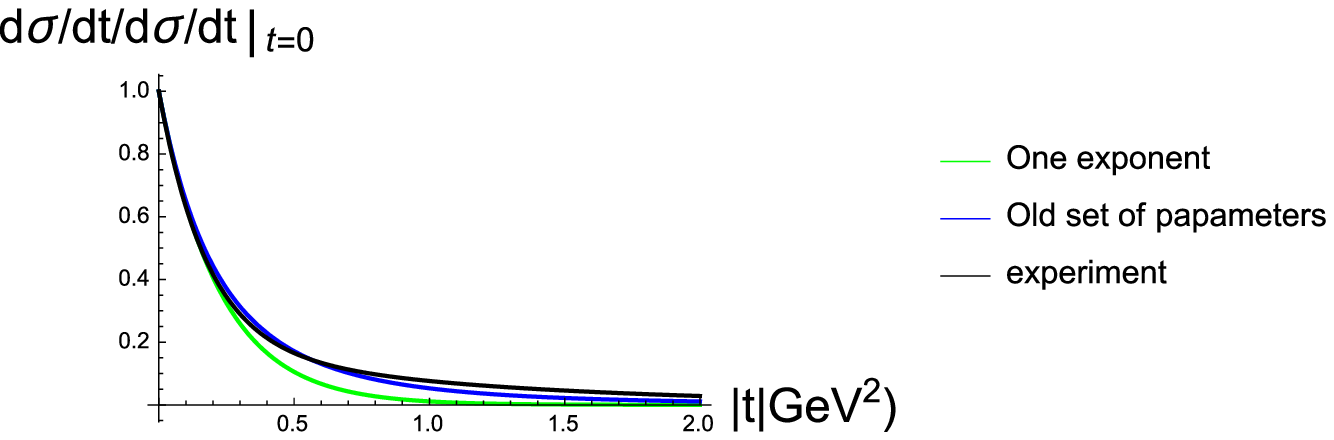} \\
      \fig{dsdt}-a\\
        \includegraphics[width=14cm]{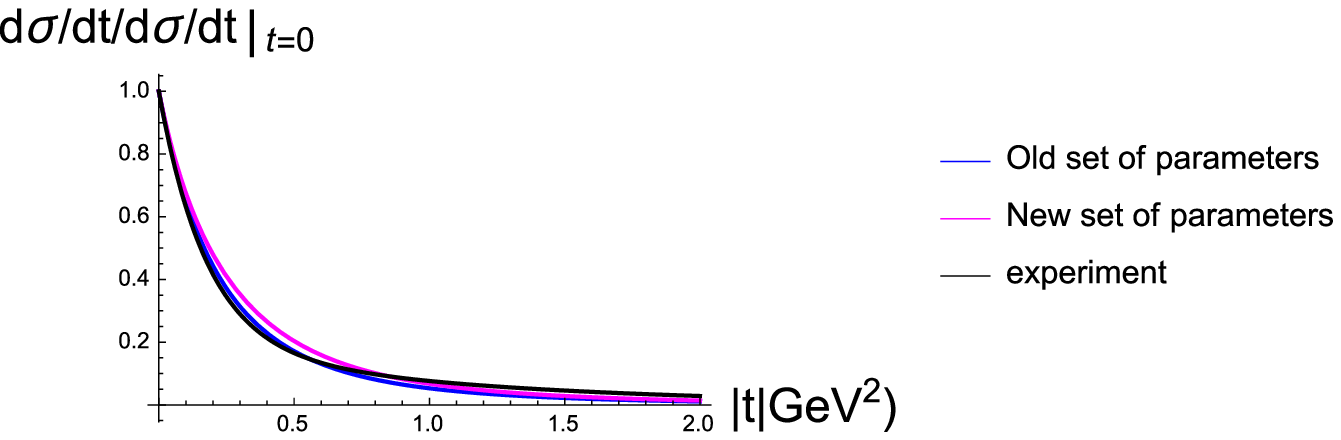} \\
        \fig{dsdt}-b\\
        \end{tabular}      
      \caption{$d \sigma/dt \Big{/}\Lb d \sigma/d t |_{t=0 }\Rb$ versus
 $|t|$: in \fig {dsdt}-a for  \eq{HARDAM}( blue,`one exponents' curve),  \eq{NIMVP}  for hard amplitude in our model, and black curve which is the fit to the experimental data taken from Ref. \cite{LHCB}; and in \fig {dsdt}-b  blue curve corresponds to the old set of the parameters in our model while  the red line describes the prediction of our model with the new set of the parameters.
   The typical experimental errors are $\pm 0.025$.}
\label{dsdt}
   \end{figure}

        We can illustrate  the importance of the $b$-dependence of the 
hard amplitude by introducing 
  \bea \label{HARDAMEX}
  S^h_{1}\,\,&=&\,\,\alpha \frac{1}{2 \pi B^h} e^{ - \frac{b^2}{2 B^h}};\nn\\
    S^h_{2}\,\,&=&\,\,\beta \frac{1}{2 \pi B^h} e^{ - \frac{b^2}{2 B^h}};  
    \eea
 with  $B^h = 4 \div 5\,GeV^{-2}$, which follows from the
 experimental data, as discussed previously.
 At first sight \eq{HARDAMEX} follows from the experimental
 observation of the vector meson production in deep inelastic
 scattering. As we have discussed, our hard amplitude of \eq{HARDAM}
 leads to the slope of the differential cross section which is the
 same as in \eq{HARDAMEX}. Indeed, as shown in \fig{dsdt},
  the $t$-dependence of the differential cross sections
in the region
 of small $t$ ($t \,<\,0.5 \,GeV^2$), 
are
 similar in both parametrizations of the hard amplitude. 
  However, at large $t$
 there is a difference, which increases with increasing $t$.
    %%%%%%%%%%%%%%%%%%%%%%%%%%%%%%%%%%%%%%%%%%%%

 %%%%%%%%%%%%%%%%%%%%%%%%%%%%%%%%%%%%%%%%%%%%%% %% 
 
\begin{figure}[ht]
    \centering
  \leavevmode
      \includegraphics[width=14cm]{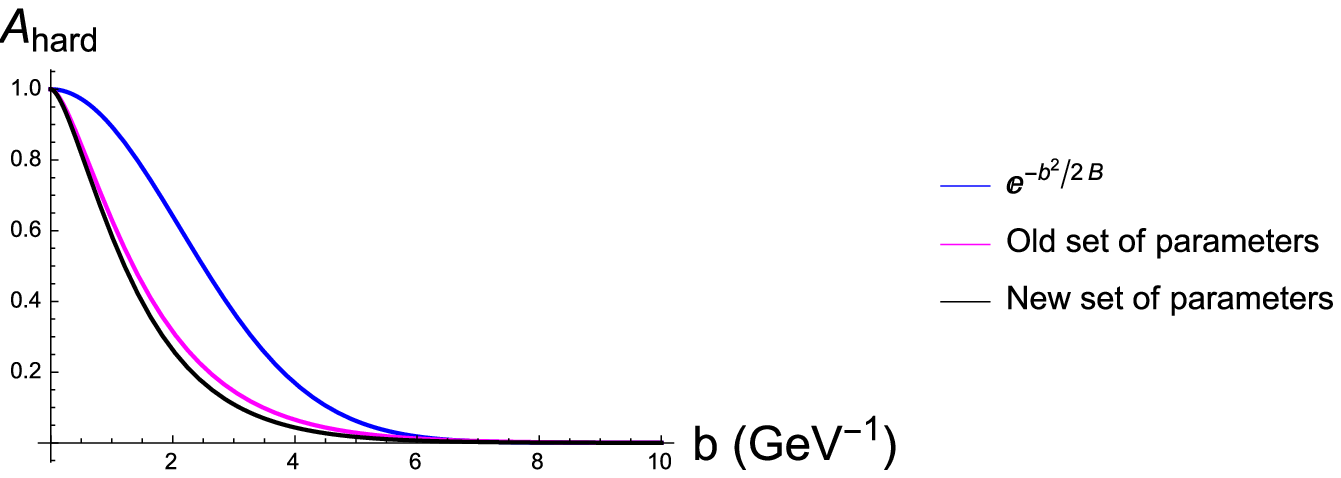}  
      \caption{ Comparison of hard amplitudes defined in
 \eq{HARDAM} and \eq{HARDAMEX}.}
\label{hardb}
   \end{figure}

 The difference between the amplitudes is more pronounced when plotted as 
a function of the impact parameter
  $b$ (see \fig{hardb}).
 
     %%%%%%%%%%%%%%%%%%%%%%%%%%%%%%%%%%%%%%%%%%%%

 %%% %%%%%%%%%%%%%%%%%%%%%%%%%%%%%%%%%%%%%%%%%%
  Figure \fig{hardb}
  shows that \eq{HARDAM} of our model (with the old set of parameters) 
leads to fast 
decrease
 of the amplitude in $b$.
  As one can see from Table 1, the steepest decrease is due to
 the $A_{2 2}(b)$ amplitude, which is the smallest. This amplitude
also  has
  the smallest suppression, due to the  factors $\exp\Lb -
 2 \Omega_{i k }\Rb$, since $\Omega_{2 2}$ has the smallest value.
  In the numerator of \eq{FF1}, only term $( 2 2)$
 provides the  essential contribution.  This term turns out to be rather
 large for the amplitude of \eq{HARDAMEX}, as one can see from 
\fig{hardb}.
 On the other hand, the same hard amplitude in \eq{HARDAM} shows the steep
 decrease in $b$, resulting in the small contribution to the numerator of
 \eq{FF1}, as well as for the resulting $\langle S^2 \rangle$.

 The resulting difference for $\langle S^2 \rangle$ is
 large,  values of the survival probability for the hard
amplitude of \eq{HARDAMEX}, are ten or more, times larger than the
results of our present model (with the old set of parameters).
 For example, for $W = 7 \,TeV$ we obtain $\langle S^2\rangle\, = \,10
 \div 15\%$.

 We denote our fit which results in these very low 
values of $\langle S^2 \rangle$ as the         "old set of parameters".
Based on the our diagnosis of the problem above, we made a second fit to 
t he same experimental data, with the additional condition that $m_{2} 
\leq$ 1.5 $GeV$. We 
will refer to this fit as "new set of parameters". The values of the 
parameters for both "old" and "new" sets are given in Table 1.
The comparison of the results for  $\sigma_{tot}, \sigma_{el}, B_{el}$,
$\sigma_{sd}$(low and high mass) and $\sigma_{dd}$ (low and high mass) for 
both set of parameters are shown in Table 4. We note that the values of
$\sigma_{tot}, \sigma_{el}$ and $B_{el}$ obtained in both fits are rather 
close, while the diffractive cross sections, both $\sigma_{sd}$ and 
$\sigma_{dd}$ are smaller in the new fit.

The results for $\langle S^2 \rangle$ for the "new set of parameters" is 
shown in Fig.12, we find \\ $\langle S^2 \rangle \approx$
3 \% in the LHC energy range.

 %%%%%%%%%%%%%%%%%%%%%%%%%%%%%%%%%%%%%%%%%%%%%%%%%%
 \subsection{Kinematic  corrections}
 %%%%%%%%%%%%%%%%%%%%%%%%%%%%%%%%%%%%%%%%%%%%%%%%%%

  In our approach we consider $G_{3\pom}$ as a point-like vertex.
 This assumption is  a considerable simplification. As
 we have discussed in \cite{GLMCOR},  we expect  short range
 correlations in rapidity, with the correlation length in rapidity
 $\Delta_{cor} \approx 2 $.  Bearing in mind that the triple Pomeron
 vertex has a size in rapidity, we can take into consideration that
 in \eq{ENH1} and \eq{OMG1}, the Pomerons enter not at rapidity $Y$
 but at $Y - \delta_{cor} N_{G_{3\pom}}$, where $N_{G_{3\pom}}$ is
 the average number of  triple Pomeron vertices . It is easy to
 see that
  \beq \label{NG}
  N_{G_{3\pom}}\,\,=\,\,\int d^2 b  T\Lb Y, b \Rb  \frac{d   
 G^{\mbox{\tiny dressed}}\Lb T\Lb Y, b \Rb\Rb}{ T\Lb Y, b \Rb}\Bigg{/} 
  \int d^2b\, G^{\mbox{\tiny dressed}}\Lb T\Lb Y, b \Rb\Rb . 
  \eeq
  
\begin{figure}[ht]
    \centering
  \leavevmode
      \includegraphics[width=14cm]{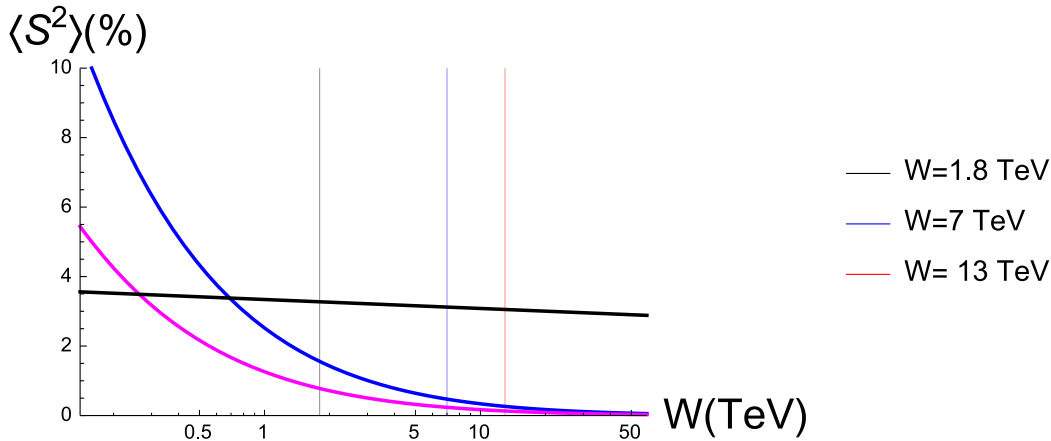}  
      \caption{ $\langle S^2\rangle$ versus $W$. $\langle S^2\rangle$
 is calculated using \protect\eq{FF1} and \protect\eq{FF2} with $Y  $
 replaced by $ Y - \Delta_{cor} \,  N_{G_{3\pom}}$ with $\Delta_{cor}
 =2$ ( blue curve), while the red curve is the same as in \protect\fig{fin}. The black curve is the estimates for the survival probability with new set of parameters (see Table 1).}
          \label{fin1}
   \end{figure}

  In \fig{fin1} we plot $\langle S^2 \rangle$ which is given by
 \eq{FF1} and \eq{FF2} but $Y \to Y - \Delta_{cor} 
  N_{G_{3\pom}}$ with $N_{G_{3\pom}}$ estimated using \eq{NG}.
 One can see that the effect is sizeable, and leads to larger values
 of the survival probability.
  
    %%%%%%%%%%%%%%%%%%%%%%%%%%%%%%%%%%%%%%%%%%%%%% %%
              %%%%%%%%%%%%%%%%%%%%%%%%%%%%%%%%%%%%%%%%%%%%

 %%%%%%%%%%%%%%%%%%%%%%%%%%%%%%%%%%%%%%%%%%%%%% %% 

 %%%%%%%%%%%%%%%%%%%%%%%%%%%%%%%%%%%%%%%%%%%%%%%%%%
 \subsection{Comparison with other estimates}
 %%%%%%%%%%%%%%%%%%%%%%%%%%%%%%%%%%%%%%%%%%%%%%%%%% 
    Nearly ten years ago we summarized the situation regarding the
evaluation of the $\langle S^2\rangle$ by different models \cite{GLMNP}.
Unfortunately, as the results of this paper illustrate, the values 
obtained 
for $\langle S^2 \rangle$, are highly dependent on the characteristics   
of 
the 
models used to parametrize the soft and hard amplitudes.
 For  details of the parametrizations used by the three groups quoted in 
Table 2, we refer 
the reader to \cite{KKMR,PLS,GLMNP}. Their values for the $\langle S^2 
\rangle$  are given 
in Table 2.
\begin{table}
 \centering
\begin{tabular}{|l|l|l|l|}
\hline
W (GeV) & KKMR (CD) \cite{KKMR}  & Pythia \cite{PLS}   &GLM \cite{GLMNP} 
\\ 
\hline
540  & 6. & n/a & 6.6  \\
\hline
1800 & 4.5  & 4.0 & 5.5  \\
\hline
14000 & 2.  &2.6 & 3.6   \\
\hline
\end{tabular}
\caption {Values of  $\langle S^2 \rangle$ at different energies from the 
three different groups \cite{KKMR,PLS,GLMNP}. The 
results are given as  percentages.  
 }
\label{sp2005}
\end{table}
 
In the summary of our previous approach for constructing a 
model based on N =4 SYM for strong coupling, and matching with 
the perturbative QCD approach \cite{GLMREV}, we discuss the results for 
$\langle S^2 \rangle$ obtained from this approach. We compared our results 
with those obtained by the Durham group \cite{KMR13}. The Durham model is 
a two-channel eikonal model where the Pomeron coupling to the diffractive 
eigenstates are energy dependent. They presented four different versions,
in Table 3,  we quote their results for model 4, their 
"favoured version".
  
\begin{table}
 \centering
\begin{tabular}{|l|l|l|l|l|l|l|}
\hline
W (GeV) &This model(I)&This model(II) &This model(IIn)&This model(III)& KMR  \cite{KMR13} 
   &GLM \cite{GLMREV}\\
\hline
1800 & 7.6 & 0.86 &3.34& 1.68& 2.8  & 7.02   \\
\hline
7000 &3.63&0.3&3.1&0.63& 1.5  &2.98    \\
\hline
14000 &2.3 & 0.25&3.05&0.44& 1.  &1.75   \\
\hline
\end{tabular}
\caption{Values of  $\langle S^2 \rangle$ at different energies from the
 different groups \cite{KMR13} and \cite{GLMREV}  and from this model: I
 - the hard amplitude is given by \eq{HARDAM} with $B = 4.5 GeV^{-2}$ 
(see \fig{dsdt} and \fig{hardb} ),II-  the hard amplitude is given by 
\eq{NIMVP} with old set of parameters; IIn- is the same as II but with the new set of parameters;  and III- is the same as II but the kinematic corrections
 are include.The $\langle S^2 \rangle$ 
results are  in percentages.  }
\label{sp2013}
\end{table}

 As can be seen from Table 3 , our results for $\langle S^2 
\rangle$ obtained from the N =4 SYM approach are slightly larger than those 
given 
by the KMR approach, and  larger than the results of our present 
model. The reasons for this have been discussed in subsection 4.2.

 Comparing the results of different calculations given in Tables
  2 and   3 with our present calculation, we see that the estimates
 using the same $b$-dependence of the hard amplitude $A_{\mbox{hard}}
 \propto \exp\Lb - b^2/4 B\Rb$ ( see Table 3  `this model(I)') , leads
 to  results that are similar to  the estimates obtained by the other 
groups. 
 The $b$-dependence of the hard amplitude that follows from our present 
approach,
 produces  small values for the survival probabilities with the 
 old set of parameters, and reasonable values with the new set of 
parameters.  We wish
  to emphasize that our b -dependence of our present model, has two 
advantages: it leads
 to the correct Froissart limit at large $b$, \, $A_{\mbox{hard}} \propto
 \exp\Lb - \mu\,b\Rb$; and at large momentum transfer ($Q_T$) it
 decreases as a  power of $Q_T$, as one expects in perturbative QCD.
 
 We wish to emphasis, that we made  a second (new) fit to the experimental 
data, and obtained a new set of parameters,
 which  leads to an increase in the values of the survival probabilities
 as  shown in Table 3.

%%%%%%%%%%%%%%%%%%%%%%%%%%%%%%%%%%%%%%%%%%%%%%%
\section{Conclusions}
%%%%%%%%%%%%%%%%%%%%%%%%%%%%%%%%%%%%%%%%%%%%%
 In this paper we calculated the survival probability for central
 diffractive production, and found at LHC energies, that its value is 
small. 
  The small value obtained does not stem from the sum of 
enhanced
 diagrams, as  in our previous models\cite{GLMSP3}, but  
is due to
 the impact parameter dependence of the hard amplitude.
 
   The distinguishing feature of our model based on CGC/saturation 
approach, is that we use  a 
framework where soft and hard processes are treated on the same  footing. 
Hence,
 there is no  need to introduce a special hard amplitude,
 as has been done in all previous attempts, to estimate the survival
 probability. It should be stressed that the main source for our small
values 
 of $\langle S^2 \rangle$, is the impact parameter dependence 
of the hard
 amplitude, for which we do not have any theoretical estimate. This is 
usually assumed to have a 
  Gaussian form $A_{\mbox{hard}} \,\propto\,\exp\Lb
 - b^2/\Lb 2 B\Rb\Rb$. The value of $B$ was taken from the experimental
 data on the deep inelastic diffractive production of vector mesons.
%%%%%%%%%%%%%%%%%%%%%%%%%%%%%%%%%%%%%%%%%%%%%%%%%%%%%%%%%%%%%%%%%%%%%%%%%%%%%
 We
 demonstrated in this paper, that in spite of the fact, that our hard 
amplitude leads to
 experimental values of $B$, at small $t$, it yields a  
different 
behaviour than the Gaussian input, leading to  small values
 of $\langle S^2 \rangle$ at high energies.
 We note, that the impact parameter dependence
 of our hard amplitude satisfies two
 theoretical features that are violated in the Gaussian $b$-dependence:
 at large $b$\,, $A_{\mbox{hard}} \,\propto\,\exp\Lb
 - \mu b\Rb$,  and at large $Q_T$ it decreases as a power
 of $Q_T$, as required by perturbative QCD.
 
  We wish to stress that the values obtained for  the survival 
probability
 depend mostly on the $b$-dependence of the hard amplitude. The most
 interesting result  is that we can describe on the same footing both
 the soft and the hard amplitude. 
 At first sight, the small values of
  $\langle S^2 \rangle$  contradict this the most basic idea of our
 approach.
  To show that this is
 not  an inherent problem of our approach, we made a new fit to the all 
available soft 
data to show
 that we can obtain substantially larger values   of the survival
 probability. It demonstrates that experimental measurements of
 this observable is a sensitive tool to determine the values of
 the phenomenological parameters of our model. 
  
 We present in this paper, the result of the first 
consistent approach
 to obtain both the soft and the hard amplitude from the same model.
  We hope that the data from the LHC on
 the survival probability,
will be  instrumental in determining  the impact parameter
 dependence of the scattering amplitude.
%%%%%%%%%%%%%%%%%%%%%%%%%%%%%%%%%%%%%%%%%%%%%%%%%%%%%%%%%%%%%%%%%%%%%%%%%%%
%%%%%%%%%%%%%%%%%%%%%%%%%%%%%%%%%%%%%%%%%%%%%%%%%%%%%%%%%%%%%%%%%%%%%%%%%%
%%%%%%%%%%%%%%%%%%%%%%%%%%%%%%%%%%%%%%%%%%%%%%%
\begin{center}
\begin{table}
\begin{tabular}{|l|l|l|l| l l |l l|}
\hline
W &$\sigma_{tot} $& $\sigma_{el}$(mb) &$B_{el}$&~~~single& diffraction~~ &~~~~double& diffraction~~~ \\
(TeV)   &  (mb)  &        (mb)              &          $(GeV^{-2})$& $\sigma^{LM}_{sd}$ (mb)  &$\sigma^{HM}_{sd}$ (mb)& $\sigma^{LM}_{dd}$ (mb)&$\sigma^{HM}_{dd}$ (mb)\\
\hline
0.576 &61.4(62.3)& 13 (12.9)&15.2 (15.2) &4.1( 5.64)& 1.42 (1.85)& 0.3 (0.7) &0.22 (0.46)\\
\hline
0.9 & 68.2(69.2 )&15.1 (15)&16 (16) &4.45 (6.25)& 1.89 (2.39)& 0.3 ( 0.77)&0.32 (0.67) \\
\hline
1.8& 78.2(79.2)&18.3 (18.2)&17.1(17.1)&4.87 (7.1) &2.79 (3.35) &0.28 ( 0.89)&0.55 (1.17) \\
\hline
2.74 & 82.3(85.5)&19.7 (20.2)&17.63 (17.8)&5 (7.6)&3.49 (4.07) &0.27 (0.97)&0.74 (1.62)\\
\hline
7 &99.9 (99.8)&25.6 (25)&19.6 (19.5)&5.38 (8.7)& 5.66 (6.2)&0.2(1.15)&1.46 (3.27)\\
\hline
8 & 102.1 (101.8)&26.4 (25.7)&19.8 (19.7)&5.41 (8.82)&6.03 (6.55) &0.2 (1.17)&1.68 (3.63)\\
\hline
13 & 110.6(109.3)&29.5 (28.3)&20.8 (20.6)&5.47 (9.36)&7.67 ( 8.08) & 0.17(1.27)&2.28 (5.11)\\
\hline
14 & 111.9 (110.5)&29.9 (28.7)&20.9 (20.7)&5.47 (9.44)& 7.87 (8.34) &0.17( 1.27) &2.32 (5.4)\\
\hline
57 & 137.8(131.7)&39.7 (36.2) &23.6 (23.1)&5.37 (10.85)&14.99(15.02) & 0.11(1.56) &5.86 (13.7)\\
\hline
\end{tabular}
\caption{ The values of cross sections versus
 energy. $\sigma^{LM}_{sd}$  and $\sigma^{LM}_{dd}$
 denote the cross sections for  diffraction dissociation
 in the low mass region, for single and double diffraction, which stem
 from the Good-Walker mechanism. While  $\sigma^{HM}_{sd}$  and 
$\sigma^{LM}_{dd}$
 are used for diffraction in high mass, coming from the dressed Pomeron
 contributions.}

\label{t2}
\end{table}
\end{center}

    %%%%%%%%%%%%%%%%%%%%%%%%%%%%%%%%%%%%%%%%%%
  \section{Acknowledgements}
  %%%%%%%%%%%%%%%%%%%%%%%%%%%%%%%%%%%%%%%%%%
   We thank our colleagues at Tel Aviv university and UTFSM for
 encouraging discussions. Our special thanks go to  Carlos Cantreras ,
 Alex Kovner and Misha Lublinsky for elucidating discussions on the
 subject of this paper.
   This research was supported by the BSF grant   2012124 and  by  the
 Fondecyt (Chile) grant  1140842.

   \end{document}